\documentstyle[amsmath,amssymb,graphicx]{article}

\def\be{\begin{eqnarray}}
\def\ee{\end{eqnarray}}
\def\nn{\nonumber}

\def\p{\partial}

\def\Tr{{\rm Tr}\,}

\def\l[{\phantom.[}

\hoffset= - 1.0in         

\begin{document}
\title{{\bf {On colored HOMFLY polynomials for twist knots} \vspace{.2cm}}
\author{{\bf Andrei Mironov$^{a,b,c,}$}\footnote{mironov@itep.ru; mironov@lpi.ru}, {\bf Alexei Morozov$^{b,c,}$}\thanks{morozov@itep.ru} \ and {\bf Andrey Morozov$^{b,c,d,e,}$}\thanks{Andrey.Morozov@itep.ru}}
\date{ }
}

\maketitle

\vspace{-5.0cm}

\begin{center}
\hfill FIAN/TD-13/14\\
\hfill ITEP/TH-25/14\\
\end{center}

\vspace{3.2cm}

\begin{center}
$^a$ {\small {\it Lebedev Physics Institute, Moscow 119991, Russia}}\\
$^b$ {\small {\it ITEP, Moscow 117218, Russia}}\\
$^c$ {\small {\it Moscow Physical Engineering Institute, Moscow 115409, Russia }}\\
$^d$ {\small {\it Moscow State University, Moscow 119991, Russia }}\\
$^e$ {\small {\it Laboratory of Quantum Topology,
Chelyabinsk State University, Chelyabinsk 454001, Russia }}
\end{center}

\vspace{1cm}

\begin{abstract}
Recent results of J.Gu and H.Jockers provide the lacking
initial conditions for the evolution method in the case of
the first non-trivially colored HOMFLY polynomials $H_{[21]}$
for the family of twist knots.
We describe this application of the evolution method, which finally allows one to penetrate
through the wall between (anti)symmetric
and non-rectangular representations for a whole family.
We reveal the necessary deformation of the differential expansion,
what, together with the recently suggested matrix model approach
gives new opportunities to guess what it could be for a generic
representation, at least for the family of twist knots.
\end{abstract}

\vspace{1cm}

\section{Introduction}

Calculation of colored HOMFLY polynomials
remains one of the biggest problems in modern
quantum field theory.
These are Wilson-loop averages in $3d$ Chern-Simons (CS) theory \cite{CS},
\be
H_R^{\cal L} = \ \left< \Tr_R \ P\exp\left(\oint_{\cal L}{\cal A}\right)\right>
\ee
with "colored" meaning that the gauge field ${\cal A}$ in the $P$-exponent
is in an arbitrary representation $R$ of the gauge group $SU(N)$.
Representation dependence of Wilson-loop averages is highly non-trivial
and very informative: for example, in confinement phase of QCD
the average in the fundamental representation obeys the area law,
while in the adjoint one it is just the perimeter law.
Chern-Simons theory is topological, therefore there is no room
for metric dependencies (like area or perimeter laws),
instead the averages depend in a sophisticated way on topology (linking)
of the closed contour ${\cal L}$ (which is called knot or link depending
on the number of connected components).

Amusingly, besides the direct analogy,
there is also a transcendental relation to Yang-Mills theory in
higher dimensions: knot polynomials are made from the Racah matrices of
quantum groups, which define the modular transformation of $2d$ conformal blocks,
and these are related through the celebrated AGT relations \cite{AGT,AGT1,AGT2}
to $S$-dualities between various $4d$ and $5d$ supersymmetric Yang-Mills models.

A dream-like solution to the colored HOMFLY problem is known for the
special class of torus knots ${\cal L}=[m,n]$:
they are provided by the action of the simplest cut-and-join operator
\be
\hat W_{[2]} = \frac{1}{2}\sum_{a,b=1}^\infty
\left((a+b)p_ap_b\frac{\p}{\p p_{a+b}} + abp_{a+b}\frac{\p^2}{\p p_a\p p_b}\right)
\ee
on the Adams-transformed characters (Schur functions) $\chi_R\{p_k\}$:
\be
H_R^{[m,n]} = \left.q^{\frac{n}{m}\hat W_{[2]}}  \chi_R\{p_{mk}\}\right|_{p_k =
p_k^*=\frac{\{A^k\}}{\{q^k\}}}
\label{RJ}
\ee
Here $q=\exp\left(\frac{2\pi i}{g+N}\right)$ is made from the gauge coupling $g$,
parameter $A=q^N$,
and we use the standard notation $\{x\}=x-x^{-1}$, so that the quantum number
is $[x]=\frac{\{q^x\}}{\{q\}}$.
A remarkable fact about the Wilson averages in Chern-Simons theory in the simply connected space-time
$R^3$ or $S^3$ is that they are {\it polynomials} in $q$ and $A$, see \cite{Witint}
for the latest attempt to explain this remarkable property. The
Adams transformation is the relabeling of time-variables
$p_k \longrightarrow p_{mk}$.

This Rosso-Jones formula (\ref{RJ}), which we presented in the version of \cite{DMMSS},
is valid in this form for the torus {\it knots}, i.e. when $m$ and $n$ are co-prime.
For the $l$-component links ($l$ is the biggest common divisor of $m$ and $n$)
the single character is substituted by the product of $l$ different characters
(which can all be in different representations).
This formula allows several important reformulations:
in terms of TBEM eigenvalue matrix model \cite{TBEM},
in terms of evolution along the "time" $n$ \cite{DMMSS,evo}
and in terms of differential expansion of \cite{DGR,arth}.
It also allows a straightforward deformation to superpolynomials
\cite{DMMSS,AgSha,Che}, and, along the lines of \cite{DM3},
perhaps, also to Khovanov-Rozansky polynomials (where the answers
are known from an alternative approach of \cite{Gortor}).

Of course, eq.(\ref{RJ}) is extremely inspiring:
one can think about writing something similar for other knots,
by making use of a similar continuation from the topological locus
$p_k^*$ to arbitrary times $p_k$ \cite{knMMM1}
and extending $\hat W_{[2]}$ evolution to that, generated by other
cut-and-join operators from \cite{MMN1}.
This Hurwitz-$\tau$-function \cite{AlMMN1} description of knot polynomials
looks promising, but only the first attempts were made \cite{Sle}.
The main obstacle is the lack of explicit examples.
An available calculation tool is essentially the old Reshetikhin-Turaev approach
\cite{RT}, either in its more traditional form of skein relations
and Racah calculus, developed very far in \cite{inds,Gu,Deg},
or in the modernized version \cite{knMMM2,IMMM,Anopaths}, based on use of the
universal ${\cal R}$-matrices and often supplemented by the old
cabling method in the new version of \cite{AnoMcabling}.
The real problem, however, is that in these tedious calculations one does not
immediately see the new structures, which are so obvious in \cite{RJ},
and which are the real target of the knot/CS theory studies.

A first real breakthrough after the old result (\ref{RJ}) was the discovery \cite{IMMMfe}
of the general formula all totally symmetric $R=[r]$ and antisymmetric $R=[1^r]$
representations for the figure-eight knot $4_1$, which was immediately
generalized to entire one-parametric family of twist knots   \cite{evo}:
\be
H_{[r]}^{(k)} = 1 + \sum_{s=1}^r  \frac{[r]!}{[s]![r-s]!}\,F_s^{(k)}(A|q)\,
\prod_{j=0}^{s-1} \underbrace{\{Aq^{r+j}\}\{Aq^{j-1}\}}_{Z_{r|1}^{(j)}}
\label{twir}
\ee
For $4_1$ ($k=-1$) all the coefficient functions are unities, $F_s^{(-1)}=1$,
for the trefoil $3_1$ $(k=1)$,
which is the only torus knot among the twist ones, $F_s^{(1)} = (-A^2)^sq^{s(s-1)}$
while for generic integer $k$
\be
F^{(k)}_s = q^{s(s-1)/2}A^s\sum\limits_{j=0}^s(-)^j\frac{[s]!}{[j]![s-j]!}\frac{\{Aq^{2j-1}\}(Aq^{j-1})^{2jk}}{\prod_{i=j-1}^{s+j-1}\{Aq^i\}}
\ee
In fact, the family of twist knots can be further extended \cite{evo}, see also \cite{genstwist}.
A very clear structure of differential expansion is seen in (\ref{twir}) -- in fact much more
transparent than in the torus case \cite{arth}, which allowed one to make further
conjecture about arbitrary rectangular representations $R=[s^r]$ \cite{GGS,arth}.
Also straightforward was generalization to superpolynomials \cite{IMMMfe,evo},
what allowed one to check a conjecture about their representation
dependence \cite{Che,anton} and to develop the theory of (super)$A$-polynomials \cite{FGS,genstwist,arthAENV},
generalizing the old story originally known in the Jones case \cite{Gar,3dAGT}.
The latest achievement coming from the study of (\ref{twir}) is an inspiring attempt to
generalize the TBEM matrix model from the torus to twist knots \cite{AlMMMmamo}.
If it was fully successful, it would provide {\it all} colored HOMFLY polynomials for twist knots
(with arbitrary Young diagram $R$), in its present form it allows one only to check
the first terms of the $\hbar$-expansion ($q=e^\hbar$), what, in fact, is not so bad.

The problem is that going beyond the (anti)symmetric and especially beyond
rectangular representations is extremely hard: the problem for particular knots
and even particular $N$ is at the border of capacity of available computers
already for $R=[21]$, nothing to say about the most interesting case of $R=[31]$.
Actually, since \cite{AnoMcabling} $H_{[21]}$ for the four different twisted knots were
known but this was not enough to apply the evolution method of \cite{evo}.
This is now possible, because in \cite{Gu} $H_{[21]}$ were calculated for three more twist knots,
so we can finally look for general formulas.
Also a partial validation is provided by the recent matrix model of \cite{AlMMMmamo}.
It is the purpose of this paper to describe the result:
it is provided by eqs.(\ref{anza21}) and (\ref{u's21})
and still needs to be converted in the differential expansion form,
generalizing eq.(\ref{exa}) to arbitrary $k$.
Even in this unfinished form,
this is a next big step after the guesswork of \cite{Ano21},
it provides a support to the ideas of that paper and puts the study
of colored knot polynomials on a more solid ground.
Also, representation $[21]$ is the first non-rectangular one,
and it is the first representation distinguishing, for example, the mutant pair of the Kinoshita–Terasaka and Conway knots
\cite{Mort}.

\section{HOMFLY for $R=[21]$ by evaluation method}

In application to twist knots,
the evolution method explained in full detail in \cite{evo}
consists of three steps.

\bigskip

1) Decompose the product of representation $R=[21]$ and its conjugate
$\bar R = \overline{[21]} = [2^{N-2}1]$ into irreducible ones:
\be
[21]]\otimes\overline{[21]} = [432^{N-4}1]\oplus [42^{N-2}]\oplus [42^{N-3}11]
\oplus [332^{N-3}]\oplus [332^{N-4}11]\oplus\ 2\times [32^{N-2}1]\  \oplus [2^N]
\ee
Note that $[32^{N-2}1]$ comes with non-trivial multiplicity ($2$),
what never happens for (anti)symmetric representations, but is the generic case
in the study of colored HOMFLY polynomials.

One can check this decomposition by summing up dimensions
\be
D_{[432^{N-4}1]}(N) = \frac{N!\frac{(N+1)!}{2\cdot 3}(N+1)(N+2)(N+3)}{3\cdot(N-4)!(N-3+1)(N-2+2)
\cdot\frac{(N-3+1)!}{2}(N-2+2)(N-1+3)}
= \frac{(N^2-9)(N^2-1)^2}{9}, \nn \\
D_{[42^{N-2}]}(N) = \frac{N!\frac{(N+1)!}{2}(N+2)(N+3)}{2\cdot(N-2+1)!(N-1+3))(N-2)!(N-1+3)}
= \frac{N^2(N-1)(N+3)}{4}, \nn \\
D_{[42^{N-3}11]}(N) = D_{[31^{N-3}]}(N) = \frac{\frac{N!}{2}(N+1)(N+2)}{2(N-3)!(N-2+2)} =
\frac{(N^2-4)(N^2-1)}{4}, \nn \\
D_{[332^{N-3}]} = \frac{N!\frac{(N+1)!}{2}(N+2)(N+1)}{2(N-3)!(N-2+1)(N-1+1)(N-3+1)!(N-2+2)(N-1+2)}
= \frac{(N^2-4)(N^2-1)}{4}, \nn \\
D_{[332^{N-4}11]} = D_{[221^{N-4}]}=
\frac{\frac{N!}{2}(N+1)N}{2(N-4)!(N-3+1)(N-2+1) } = \frac{(N+1)N^2(N-3)}{4}, \nn \\
D_{[32^{N-2}1]}(N)=D_{[21^{N-2}]} = \frac{N!(N+1)}{(N-2)!(N-1+1)} = N^2-1, \nn \\
D_{[2^N]} =D_{[0]} = 1
\ee
the sum is indeed equal to the square of $D_{[21]}(N)=\frac{N(N^2-1)}{3}$.

\bigskip

2) The $k$-dependence of $H^{(k)}$ is dictated be the
eigenvalues of $\hat W_{[2]}$, $\varkappa_R=\sum_{(i,j)\in R} (i-j) - \sum_{(i,j)\in [2^N]} (i-j) $:
\be
\varkappa_{[432^{N-4}1]} = 3+2+1+(N-1)+(N-2)+(N-3)=3N &\Longrightarrow & A^3 ,\nn \\
\varkappa_{[42^{N-2}]} =3+2+(N-1)+(N-2)=2N+2 & \Longrightarrow & q^2A^2 ,\nn \\
\varkappa_{[42^{N-3}11]} =3+2+(N-2)+(N-3)= 2N&\Longrightarrow &A^2 ,\nn \\
\varkappa_{[332^{N-3}]} =2+1+(N-1)+(N-2)=2N&\Longrightarrow &A^2 ,\nn \\
\varkappa_{[332^{N-4}11]} =2+1+((N-2)+(N-3)=2N-2&\Longrightarrow & q^{-2}A^2 ,\nn \\
\varkappa_{[32^{N-2}1]} = 2+(N-2) = N&\Longrightarrow & A ,\nn \\
\varkappa_{[2^N]} = 0 & \Longrightarrow & 1
\ee
This is because they are actually the eigenvalues of quantum ${\cal R}$-matrix,
which acts as the unit operator in each irreducible representation in the
$2$-strand channel.

In other words, one makes the following anzatz for the evolution along the $k$-variable:
\be
H^{(k)}_{[21]} = u_3A^{6k} + \Big(u_{2p}q^{4k}+u_{20}+u_{2m}q^{-4k}\Big)A^{4k} +u_1A^{2k} + u_0
\label{anza21}
\ee

\bigskip

3) The six unknown $k$-independent coefficients $u_\alpha(A,q)$ in (\ref{anza21})
can be now defined from "the initial conditions":
the actual values of $H_{[21]}^{(k)}$ for particular values of $k$.
To determine the six parameters, one needs six explicitly known answers for $H_{[21]}^{(k)}$.
Immediately available are two: for the unknot at $k=0$ and for the torus knot, trefoil at $k=1$.
Two more, for the $3$-strand knots $4_1$ at $k=-1$ and $5_2$ at $k=2$
were found by a tedious cabling calculation in \cite{AnoMcabling},
but this was not enough.
Quite recently, Jie Gu and Hans Jockers published their results \cite{Gu}
for the four-strand $6_1$ ($k=-2$) and $7_2$ ($k=3$) and even the
5-strand $8_1$ ($k=-3$): they use an alternative group theoretical approach
{\it a la} \cite{inds}, and are not restricted with the number of strands in the closed braid since use plat representations of knots.
This allows us not only to apply the evolution method,
but even provides the seventh point in the $k$-line to check the outcome.

\bigskip

The answer is:
\be
u_3   = -A^3\frac{\{Aq^3\}\{A/q^{3}\}\{Aq\}\{A/q\}}{\{A\}}
\nn \\
u_{2p}= \frac{[3]}{[2]^2} A^3\frac{\{Aq^3\}\{A\}}{\{Aq\}}
\Big([2]A^2q^{-3}-(q^4+1-q^{-2}+q^{-4})\Big) \nn \\
u_2  = -2\{q\}^2 A^3\left(\frac{[3]}{[2] }\right)^2\frac{\{Aq^2\}\{A/q^2\}}{\{A\}}
\nn\\
u_{2m}= \frac{[3]}{[2]^2}A^3\frac{\{Aq^{-3}\}\{A\}}{\{Aq^{-1}\}}
\Big([2]A^2q^3 -(q^4+1-q^{2}+q^{-4})\Big)
\nn\\
u_1 = -\frac{[3]A^3}{\{A\}}\Big(A^4 -(q^6+q^{-6})A^2 + (2q^6-4q^4+4q^2-3+4q^{-2}-4q^{-4}+2q^{-6})\Big)
\nn\\
u_0 = \frac{A^3}{\{Aq\}\{A\}\{A/q\}}\left(A^6-\frac{[3][10]}{[2][5]}A^4
+ \frac{[3][10]^2}{[2]^2[5]^2}A^2-\frac{[10][14]}{[2]^2[5][7]}\right)
\label{u's21}
\ee

\bigskip

Now, when we possess the general expression for $H_{[21]}^{(k)}$,
it is easy to check some of its crucial properties
enumerated in \cite{Ano21}.

\section{Checking elementary properties}

For $A=q^2$ the gauge group is $SL(2)$, for which there is no difference
between representations $[21]$ and $[1]$, thus (\ref{u's21})-(\ref{anza21}) coincide
with
\be
H_{[1]}^{(k)} = 1 + F_1^{(k)}\{Aq\}\{A/q\} = 1 - \frac{A^{k+1}\{A^k\}}{\{A\}}\{Aq\}\{A/q\}
\ \ \stackrel{A=q^2}{\longrightarrow} \ \
J_{[1]}^{(k)}(q) = \frac{q^2+q^6 + (1-q^6)\cdot q^{4k}}{1+q^2}
\label{Jo21}
\ee
The same coincidence takes place at $A=q^{-2}$ , this follows also from
the symmetry
\be
H^{(k)}_{[21]}(A,q^{-1}) = H^{(k)}_{[21]}(A,q)
\ee

For $A=q$ one could expect that the knot polynomial vanishes, and this is indeed
true, but for the {\it unreduced} HOMFLY polynomial.
As to (\ref{u's21})-(\ref{anza21}), it is the {\it reduced} polynomial,
and nothing special happens to it at $A=q$:
what vanishes at $A=q$ is the quantum dimension of representation $[21]$, i.e.
$\chi^*_{[21]}=\frac{\{Aq\}\{A\}\{A/q\}}{\{\{q\}^2\{q^3\}}$.

At $A=1$, one obtains the Alexander polynomial
\be
H_{[21]}^{(k)}(A=1,q) = 1 + k\{q^3\}^2 \ = \ H_{[1]}^{(k)}(A=1,q^3)
\ee
(the last relation \cite{DMMSS,IMMMfe,Zhu} holds only for {\it hook} diagrams,
but representation $[21]$ is of that type).

Finally, for $q=1$, one gets the {\it special} polynomial, and
\be
H_{[21]}^{(k)}(q=1,A)  = \left(\ A^2\cdot\Big(1-A^{2k-1}\{A\}\Big)\ \right)^3
= \Big(H_{[1]}^{(k)}(q=1,A)\Big)^3
\ee
in full accordance with \cite{DMMSS},  \cite{anton} and \cite{Zhu}.

\section{Checking consistency with the matrix model}

Another comparison to make is with the matrix model suggestion of \cite{AlMMMmamo}.
Since there are no reason for any doubt about the answer (\ref{anza21})+(\ref{u's21})
for $H^{(k)}_{[21]}$, this is rather a check of the matrix model.
However, an advantage of the matrix model approach is that its calculation
complexity is almost independent of the representation, and, once developed,
it provides generic colored polynomials.
Thus, checks and insights about this approach are extremely important
for practical calculations, not only for pure theory.

The claim of \cite{AlMMMmamo} is that the colored Jones polynomial (i.e. HOMFLY polynomial at $N=2$,
$A=q^2$) possesses a
remarkable integral representation:
\be
J_{r}(q=e^\hbar)=H_{[r-1]}(A=q^2,q=e^\hbar) \sim \int e^{-\frac{u^2}{2\gamma\hbar}}\, \sinh (ru)\,
\underbrace{e^{-\frac{\gamma\hbar}{2}\p_u^2}
{\cal J}\left(\frac{u}{\gamma}\Big|\,\hbar\right)}_{\nu(u)}
\label{intJones}
\ee
where ${\cal J}(\rho|\hbar) = J_{r=\rho/\hbar}(q=e^\hbar)$
is the same Jones polynomial, only with variables changed to describe the vicinity of the
large-representation (Kashaev) limit.
Explicit expressions for ${\cal J}$ are difficult to get, even if {\it some}  formulas
(like hypergeometric series of \cite{Gar}) are known for generic colored Jones polynomials,
but differential expansion like (\ref{twir}) is exactly what is needed for this purpose,

An immediate lift of (\ref{intJones}) to arbitrary $N$,
\be
H_R\Big(q=q^\hbar,A=e^{\hbar N}\Big) \ \stackrel{?}{\sim} \
\int \chi_R[e^u] \prod_{i<j}^N \Big(\nu(u_i-u_j)\cdot\sinh(u_i-u_j)\Big)\prod_{i=1}^N
e^{-\frac{u_i^2}{\gamma\hbar}}\,du_i
\label{intHOM}
\ee
is known \cite{TBEM} to give the colored HOMFLY polynomials
for the torus knots, but can not do so for generic knots, because
the HOMFLY polynomial can not be reconstructed from the Jones one in a knot-independent way.
Already for the twisted knots there are corrections \cite{AlMMMmamo} to (\ref{intHOM}),
but they start from the order $\hbar^5$ and seem to have a controllable dependence
on $R$ and $k$, which is currently under investigation.
At present, one can use this technique to find the $\hbar$ expansion of $H_R$
up to the terms $\hbar^6$ and for $N\leq 5$.

We performed this check and made sure that (\ref{anza21})+(\ref{u's21})
are in this sense consistent with \cite{AlMMMmamo}, what is not at all trivial,
because the input in \cite{AlMMMmamo} is only from knowledge of the HOMFLY polynomials
in symmetric representations.
In particular, we confirmed, that the first non-vanishing correction to (\ref{intHOM})
is given by the factor (38)  of \cite{AlMMMmamo} with $r$ substituted by $|R|$
(the number of boxes in $R$, i.e. $3$ for $R=[21]$):
\be
1\ +\ 2(N-2)(3N-4)\Big(2N\varkappa_R + |R|(N^2-|R|)\Big)U^{(k)}\hbar^5 + O(\hbar^6),\nn \\
U^{(k)} = k(k+1)(4k-1) + \frac{48k^2}{\gamma^3}(39k^2-13k+1)
\label{UUU}
\ee
The parameter $\gamma$ (which is equal to $-mn$ in the matrix model of \cite{TBEM}
for the $[m,n]$ torus knot, e.g. $U^{(1)}(\gamma=-6)=0$ for the trefoil)
remains unspecified: all formulas of \cite{AlMMMmamo} hold for arbitrary
value of $\gamma$, which can still be adjusted, perhaps, with some other
free parameters of similar type, to get rid of corrections like (\ref{UUU}).

\section{On differential expansion for $H_{[21]}^{(k)}$}

Despite all these successes with eqs.(\ref{anza21})+(\ref{u's21}),
they are still far from looking like the basic formula (\ref{twir}),
i.e. are not yet represented in the desired form of the differential expansion
{\it a la} \cite{IMMMfe,evo} and \cite{arth},
which is a $q$-deformation of the binomial expansion for the {\it special} polynomial
\be
H^{(k)}_R(q=1,A) = \Big(H^{(k)}_{_\Box}(q=1,A)\Big)^{|R|} =
\Big( 1   -  (A^{2k}-1)(A^2-1)\Big)^{|R|}
= \sum_{j=0}^{|R|} C^{\,j}_{|R|}
\Big( -  (A^{2k}-1)(A^2-1)  \Big)^j
\ee
From \cite{Ano21} we know that the answer should be rewritten in the form
\be
H^{(k)} = 1 +\Big(Z_{2|0}+Z_{3|3}+Z_{0|2}\Big)\cdot F_1^{(k)}(A) + \{A\}\cdot Z_{2|2}\cdot G^{(k)}(A|q)
\ee
where
\be
F_1^{(k)} = -A^2\frac{A^{2k}-1}{A^2-1}
\ee
does not depend on $q$,

Nice representations of this type are known for $k=\pm 1$:
$$
H^{(-1)}_{[21]} = 1 + \Big(Z_{2|0}+Z_{3|3}+Z_{0|2}\Big)
+ Z_{2|2}\Big(Z_{4|0}+Z_{0|4}+Z_{0|0}
\Big)
+ Z_{3|3}Z_{2|2}Z_{0|0}
-\{q\}^2 Z_{2|2}Z_{0|0}
$$
\vspace{-0.5cm}
\be
H^{(-1)}_{[21]} = 1
\label{exa}
\ee
$$
H^{(1)}_{[21]} = 1 -A^2\Big( Z_{2|0}+Z_{3|3}+Z_{0|2}\Big)
+ A^4 Z_{2|2}\Big(q^3Z_{3|0}+q^{-3}Z_{0|3} +Z_{0|0}
\Big)
-A^6 Z_{3|3}Z_{2|2} Z_{0|0}
+ A^4(1+A^2)\{q\}^2  Z_{2|2}Z_{0|0}
$$
However, what should be the right representation for $G^{(k)}(A|q)$
for generic $k$
remains unclear, this adds to the problems with the choice of
the differential expansion for the torus knots reported in \cite{arth}.
The situation is unclear even at the level of (reduced)
Alexander polynomial: at $A=1$, one has
\be
G^{(k)}(A=1) = \sum_{i=0}^{2k}  [4k+1-2i]\cdot u_i^{(k)}
\ee
where the coefficients are almost independent of $k$,
but look somewhat ugly.
\be
u_i^{(k)} = u_i^{(k-1)} \ \ \ {\rm for} \ \ \ i\leq 2k-4, \nn \\
u_{2k-3}^{(k)} = -\frac{1}{6}k(4k^2-21k+23)   \ \ {\rm for}\ k>1, \nn\\
u_{2k-2}^{(k)} = 8+\frac{1}{6}(k-2)(4k^2-7k+45) \ \ {\rm for}\ k>1, \ \ u_0^{(1)}=1, \nn\\
u_{2k-1}^{(k)} =2 -\frac{1}{6}(k+1)(4k^2-19k+18),\nn\\
u_{2k}^{(k)} = 1 + \frac{1}{6}(k-1)(4k^2-5k+18)
\ee
We plan to return to discussion of different options here
in a separate publication.

\section*{Acknowledgements}

We are indebted to Jie Gu and Hans Jockers for important
checks and additional efforts to enlarge the list
of calculated quantities and to Satoshi Nawata for pointing
out a misprint in our main formula (10). 
Our work is partly supported by grant NSh-1500.2014.2, by RFBR grants 13-02-00457 (A.Mir.), 13-02-00478 (Al.Mor.), 14-02-00627 (And.Mor.), 14-01-31395\_young\_a (And.Mor.), by joint grants 13-02-91371-ST and 14-01-92691-Ind, by the Brazil National Counsel of Scientific and Technological Development (Al.Mor.), by the Laboratory of Quantum Topology of Chelyabinsk State University (Russian Federation government grant 14.Z50.31.0020) (And.Mor.) and by the Dynasty Foundation (And.Mor.).

\end{document}